\documentclass{article} 
\usepackage{iclr2026_conference,times}

\usepackage{amsmath,amsfonts,bm}

\def\eqref#1{equation~\ref{#1}}

\def\1{\bm{1}}

\DeclareMathAlphabet{\mathsfit}{\encodingdefault}{\sfdefault}{m}{sl}
\SetMathAlphabet{\mathsfit}{bold}{\encodingdefault}{\sfdefault}{bx}{n}

\usepackage{hyperref}
\usepackage{url}
\usepackage{algorithm}
\usepackage{algorithmic}
\usepackage{subcaption}
\usepackage{booktabs}
\usepackage{graphicx}
\usepackage{adjustbox}
\usepackage{amsmath}
\usepackage{amssymb}
\usepackage{tcolorbox}

\title{Don't Retrieve, Generate: Prompting LLMs for Synthetic Training Data in Dense Retrieval}

\author{Aarush Sinha \\
Depatment of Computer Science, University of Copenhagen \\
\texttt{aarush.sinha@gmail.com}
}

\iclrfinalcopy 
\begin{document}

\maketitle

\begin{abstract}
Training effective dense retrieval models typically relies on hard negative (HN) examples mined from large document corpora using methods such as BM25 or cross-encoders, which require full corpus access and expensive index construction. We propose generating synthetic hard negatives directly from a provided query and positive passage, using Large Language Models(LLMs). We fine-tune DistilBERT using synthetic negatives generated by four state-of-the-art LLMs ranging from 4B to 30B parameters (Qwen3, LLaMA3, Phi4) and evaluate performance across 10 BEIR benchmark datasets. Contrary to the prevailing assumption that stronger generative models yield better synthetic data, find that our generative pipeline consistently underperforms traditional corpus-based mining strategies (BM25 and Cross-Encoder). Furthermore, we observe that scaling the generator model does not monotonically improve retrieval performance and find that the 14B parameter model outperforms the 30B model and in some settings it is the worst performing.
\end{abstract}

\section{Introduction}

Dense retrieval models, which encode queries and documents into low-dimensional vectors, are foundational to modern information retrieval systems (\cite{karpukhin2020dense, xiong2020approximate, reimers2019sentence}). These models typically require fine-tuning on task-specific data using contrastive objectives like triplet loss, which pulls relevant pairs closer while pushing away irrelevant (negative) passages \cite{henderson2017efficient}.

A key factor in effective fine-tuning is the quality of negative examples (\cite{xiong2020approximate}). While random negatives are easy to generate, hard negatives semantically related but irrelevant passages provide stronger learning signals (\cite{zhan2021optimizing}). Traditionally, these are mined from the corpus using methods like BM25 or neural cross-encoders (CEs) for re-ranking (\cite{nogueira2020passage, qu2020rocketqa}). 

Large Language Models (\cite{brown2020language, touvron2023llama, geminiteam2023gemini}) have become exceedingly good at understanding and generating text by following user instructions. We propose an alternative, corpus-free pipeline that utilizes the generative abilities of LLMs. Our approach includes:
\begin{enumerate}
    \item \textbf{Hard Negative Generation:} The LLMs are prompted to produce five semantically challenging but irrelevant passages using 4 LLMs Qwen3-4B \& Qwen3-30B(\cite{yang2025qwen3technicalreport}), LLama3-8B (\cite{llama3modelcard}) and Phi4-14B (\cite{abdin2024phi4technicalreport}).
\end{enumerate}

We evaluate an end-to-end \textsc{LLM-HN} approach against two corpus-based baselines, \textsc{BM25-HN} and \textsc{CE-HN}, by fine-tuning a vanilla DistilBERT(\cite{sanh2020distilbertdistilledversionbert}) and measuring retrieval performance on 10 BEIR tasks(\cite{thakur2021beir}). We analyze the impact of LLM-generated hard negatives and additionally consider a naive combination of BM25-, cross-encoder, and LLM-based hard negatives for training, with all models evaluated on BEIR.

\section{Related Work}

Large neural models have been increasingly applied across the information retrieval pipeline. For query understanding, techniques like query expansion or generating hypothetical documents aim to improve retrieval effectiveness (\cite{zhu2023large, ding2024survey}). Generative models also serve as powerful re-ranking components, refining initial candidate lists through pairwise or listwise comparisons (\cite{ma2023zeroshotlistwisedocumentreranking, qin2024large}).

A significant area of research involves using generative models to create synthetic data for training dense retrievers, particularly valuable when labeled data is scarce. Approaches like InPars (\cite{bonifacio2022inpars}) generate relevant documents for given queries, while others focus on unsupervised data augmentation (\cite{meng2023augtriever}) or specialized data for conversational contexts (\cite{chen2024generalizingconversationaldenseretrieval}) or training smaller models (\cite{tamber2025drama}).

Most closely related to our work are approaches that use generative models to synthesize hard negative samples. SyNeg~(\cite{wang2024syneg}) leverages multi-attribute prompting and reflection and often combines synthetic negatives with corpus-mined ones. Other studies focus on optimizing negative sampling distributions(\cite{ma2024taskleveldistributionallyrobustoptimization}), exploiting citation networks or knowledge graphs(\cite{sinha2025bicaeffectivebiomedicaldense}), or generating negatives for specific settings such as conversational retrieval(\cite{jin2024instructor}).

\section{Methodology}

The initial corpus comprises 10,000 randomly sampled abstracts from ms-marco (\cite{bajaj2018msmarcohumangenerated}). We utilize the \texttt{Tevatron/msmarco-passage}\footnote{\url{https://huggingface.co/datasets/Tevatron/msmarco-passage}} which already contains queries and positive passages.

\paragraph{BM25 Indexing and Hard Negative Retrieval:}Following the construction of a unique passage corpus and its tokenization using NLTK(\cite{bird-loper-2004-nltk})  BM25\footnote{\url{https://pypi.org/project/rank-bm25/}} was initialized. This model indexed the tokenized corpus, computing necessary term statistics such as inverse document frequency (IDF). For each query-passage pair $(q, p^+)$ in the dataset, where $q$ is the query and $p^+$ is its ground-truth relevant passage, hard negative passages were retrieved. 

\paragraph{Hard Negative Mining using Cross-Encoders:}For each query-passage pair $(q, p^+)$ in the dataset, where $q$ is the query and $p^+$ is its ground-truth relevant passage, hard negative passages were retrieved using the \texttt{msmarco-MiniLM-L6-v3}. These models were initialized with help of sentence similarity (\cite{reimers2019sentence}). Cosine similarity was used as the scoring function, and follows the procedure as previous works (\cite{wang2021gpl}). 

\paragraph{Hard-Negatives Generation using LLMs:}The goal is to leverage the LLM's understanding and generative capabilities to create passages that appear relevant to a given query but do not actually answer it, serving as highly challenging negative examples for information retrieval tasks. We detail the inference setup in Appendix \ref{llm-config}. A structured chat prompt was engineered to guide the LLM in generating hard negative passages. The prompts consist of a system instruction and a user query template which are provided in Appendix \ref{prompts}.

\section{Experiments}

\subsection{Fine-Tuning}

We fine-tune DistilBERT(\cite{sanh2020distilbertdistilledversionbert}) for a single epoch using a 90/10 train-validation split, with early stopping applied after three consecutive evaluation steps without improvement. Model evaluation is performed every 100 training steps. We conduct separate fine-tuning runs for each set of hard negatives generated by the four LLMs, as well as for those obtained via BM25 and the cross-encoder. In addition, we perform fine-tuning on datasets constructed by naively concatenating the LLM-generated hard negatives with those produced by BM25 and the cross-encoder, resulting in a total of twelve distinct fine-tuning datasets.

All models were fine-tuned using the Multiple-Negative Ranking Loss (MNRL)(\cite{henderson2017efficient}), with \texttt{batch\_size=16}.

\subsection{Results}

\begin{table}[!t]
\centering
\scriptsize
\setlength{\tabcolsep}{3pt} 
\renewcommand{\arraystretch}{0.9} 
\begin{tabular}{l|cccccccccc|c}
\toprule
\textbf{Fine-Tuning Dataset} &
\rotatebox{60}{\textbf{FEVER}} &
\rotatebox{60}{\textbf{NFCorpus}} &
\rotatebox{60}{\textbf{SciDocs}} &
\rotatebox{60}{\textbf{SciFact}} &
\rotatebox{60}{\textbf{COVID}} &
\rotatebox{60}{\textbf{NQ}} &
\rotatebox{60}{\textbf{Climate-FEVER}} &
\rotatebox{60}{\textbf{ArguAna}} &
\rotatebox{60}{\textbf{Quora}} &
\rotatebox{60}{\textbf{FiQA}} &
\textbf{Avg} \\
\midrule

\multicolumn{12}{c}{\textbf{Baselines}} \\
\midrule
BM25 & \textbf{0.610} & 0.220 & 0.116 & 0.407 & 0.328 & 0.278 & \textbf{0.183} & 0.445 & 0.811 & 0.190 & 0.359 \\
Cross-encoder (CE) & 0.575 & 0.225 & \textbf{0.121} & \textbf{0.447} & 0.344 & \textbf{0.298} & 0.166 & \textbf{0.466} & 0.815 & \textbf{0.198} & \textbf{0.366} \\

\midrule
\multicolumn{12}{c}{\textbf{Aggregated (All LLMs)}} \\
\midrule
All LLMs
& 0.415 & \textbf{0.235} & 0.106 & 0.264 & 0.265 & 0.199 & 0.101 & 0.381 & 0.796 & 0.161 & 0.292 \\
All LLMs + BM25
& 0.542 & 0.222 & 0.114 & 0.364 & 0.345 & 0.260 & {0.149} & 0.422 & 0.818 & 0.195 & {0.343} \\
All LLMs + Cross-Encoder
& 0.397 & 0.213 & 0.112 & 0.250 & 0.318 & {0.284} & 0.063 & 0.449 & 0.816 & 0.169 & 0.307 \\
All LLMs + BM25 + Cross-Encoder
& {0.545} & 0.220 & {0.117} & 0.355 & 0.296 & 0.281 & 0.120 & {0.479} & {0.821} & 0.191 & 0.342 \\

\midrule
\multicolumn{12}{c}{\textbf{BM25 + Cross-Encoder + LLM}} \\
\midrule
BM25 + CE + LLaMA3-8B
& 0.501 & 0.202 & 0.080 & 0.329 & 0.279 & 0.184 & 0.129 & 0.295 & 0.822 & 0.192 & 0.301 \\
BM25 + CE + Phi4-14B
& 0.458 & 0.196 & 0.064 & 0.267 & 0.255 & 0.151 & 0.061 & 0.280 & 0.817 & 0.174 & 0.272 \\
BM25 + CE + Qwen-4B
& 0.446 & 0.190 & 0.083 & 0.276 & 0.254 & 0.160 & 0.096 & 0.215 & 0.818 & 0.180 & 0.272 \\
BM25 + CE + Qwen3-30B
& 0.326 & 0.192 & 0.075 & 0.302 & 0.286 & 0.147 & 0.067 & 0.325 & 0.813 & 0.185 & 0.272 \\

\midrule
\multicolumn{12}{c}{\textbf{BM25 + LLM}} \\
\midrule
BM25 + Phi4-14B & 0.525 & {0.231} & 0.109 & 0.387 & 0.275 & 0.184 & 0.144 & 0.409 & 0.800 & 0.161 & 0.343 \\
BM25 + LLaMA3-8B & 0.518 & 0.221 & 0.102 & 0.413 & 0.311 & 0.232 & {0.172} & 0.400 & 0.814 & 0.173 & 0.336 \\
BM25 + Qwen-4B & 0.577 & 0.212 & 0.106 & 0.367 & 0.306 & 0.270 & 0.119 & 0.421 & 0.815 & 0.182 & 0.338 \\
BM25 + Qwen-30B & 0.542 & 0.207 & 0.099 & 0.321 & 0.307 & 0.255 & 0.085 & 0.406 & 0.813 & 0.165 & 0.320 \\

\midrule
\multicolumn{12}{c}{\textbf{Cross-Encoder (CE) + LLM}} \\
\midrule
CE + Phi4-14B & 0.535 & 0.221 & 0.109 & 0.421 & \textbf{0.357} & 0.293 & 0.150 & 0.453 & \textbf{0.823} & 0.190 & 0.355 \\
CE + LLaMA3-8B & 0.555 & 0.222 & 0.105 & 0.424 & 0.301 & 0.296 & 0.158 & 0.403 & 0.819 & 0.174 & 0.346 \\
CE + Qwen-4B & 0.470 & 0.203 & 0.109 & 0.330 & 0.269 & 0.296 & 0.089 & 0.439 & 0.815 & 0.177 & 0.320 \\
CE + Qwen3-30B & 0.527 & 0.211 & 0.103 & 0.349 & 0.297 & 0.278 & 0.099 & 0.422 & 0.812 & 0.171 & 0.327 \\

\midrule
\multicolumn{12}{c}{\textbf{Standalone LLMs}} \\
\midrule
Phi4-14B & 0.439 & 0.230 & 0.109 & 0.387 & 0.275 & 0.184 & 0.144 & 0.409 & 0.800 & 0.161 & 0.314 \\
LLaMA3-8B & 0.214 & 0.200 & 0.056 & 0.331 & 0.277 & 0.102 & 0.080 & 0.349 & 0.793 & 0.068 & 0.247 \\
Qwen-4B & 0.251 & 0.223 & 0.118 & 0.237 & 0.328 & 0.211 & 0.091 & 0.414 & 0.795 & 0.148 & 0.282 \\
Qwen-30B & 0.317 & 0.217 & 0.096 & 0.377 & 0.262 & 0.160 & 0.124 & 0.398 & 0.798 & 0.145 & 0.290 \\

\bottomrule
\end{tabular}
\caption{NDCG@10 results across BEIR datasets on fine-tuning DistilBERT. \textbf{Bold} denotes the best score.}
\label{tab:beir_results}
\end{table}

\paragraph{Overall Performance:}We evaluate retrieval performance on 10 BEIR \cite{thakur2021beir} datasets using nDCG@10. As shown in Table~\ref{tab:beir_results}, classical retrieval baselines outperform all LLM-only settings. BM25 achieves an average nDCG@10 of \textbf{0.359}, while the cross-encoder baseline performs best overall with an average score of \textbf{0.366}. In contrast, models fine-tuned solely with LLM-generated hard negatives achieve substantially lower performance, with an average nDCG@10 of \textbf{0.292}, indicating that synthetic negatives alone are insufficient to match traditional retrieval supervision.

\begin{figure}[htbp]
    \centering
    \begin{subfigure}{0.48\linewidth}
        \centering
        \includegraphics[width=\linewidth]{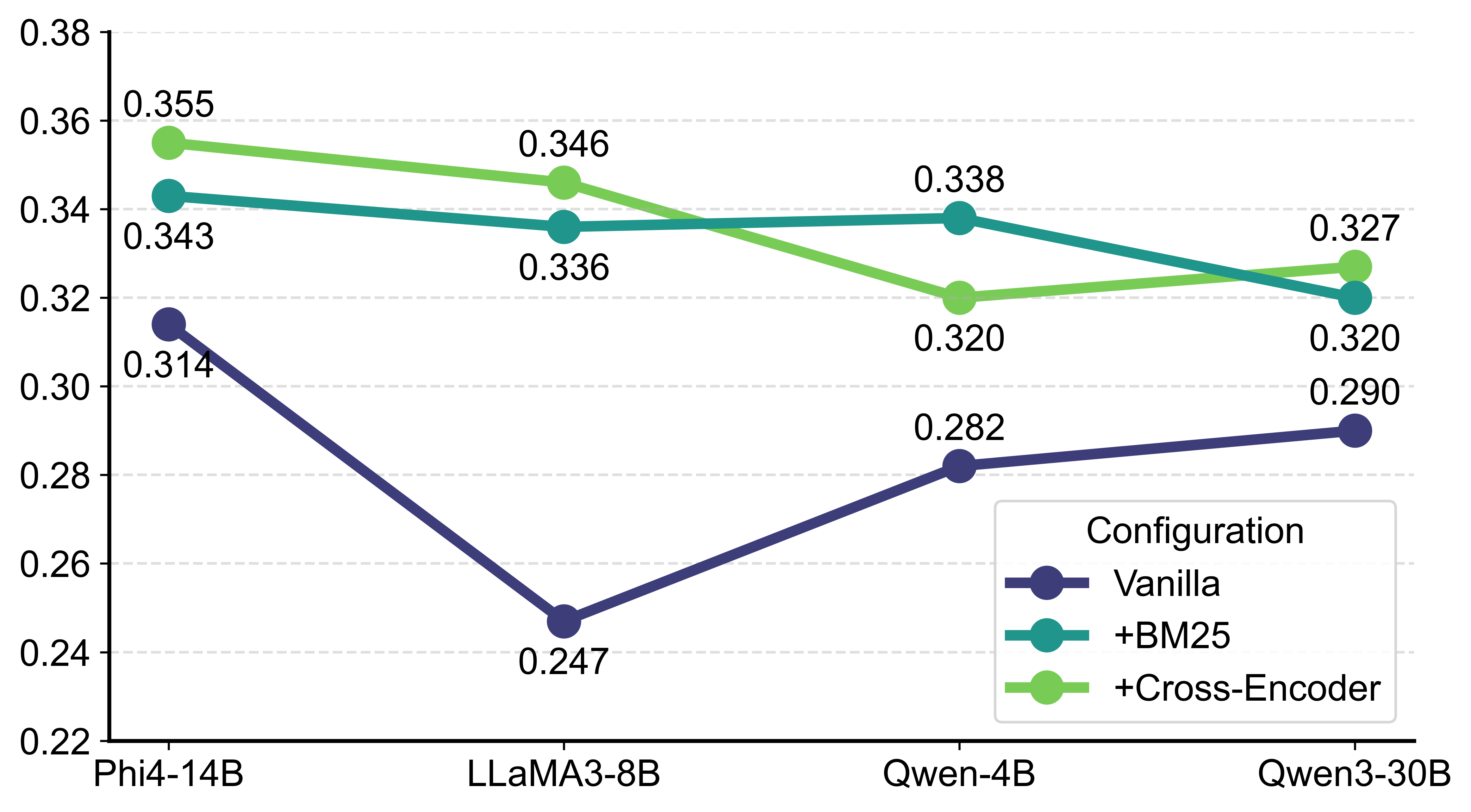}
        \caption{Average nDCG@10 on the 10 BEIR datasets for DistilBERT fine-tuned using LLM-generated hard negatives, versus fine-tuning with LLM hard negatives combined with BM25- or cross-encoder-mined hard negatives.}
        \label{fig:line_plot}
    \end{subfigure}
    \hfill
    \begin{subfigure}{0.48\linewidth}
        \centering
        \includegraphics[width=\linewidth]{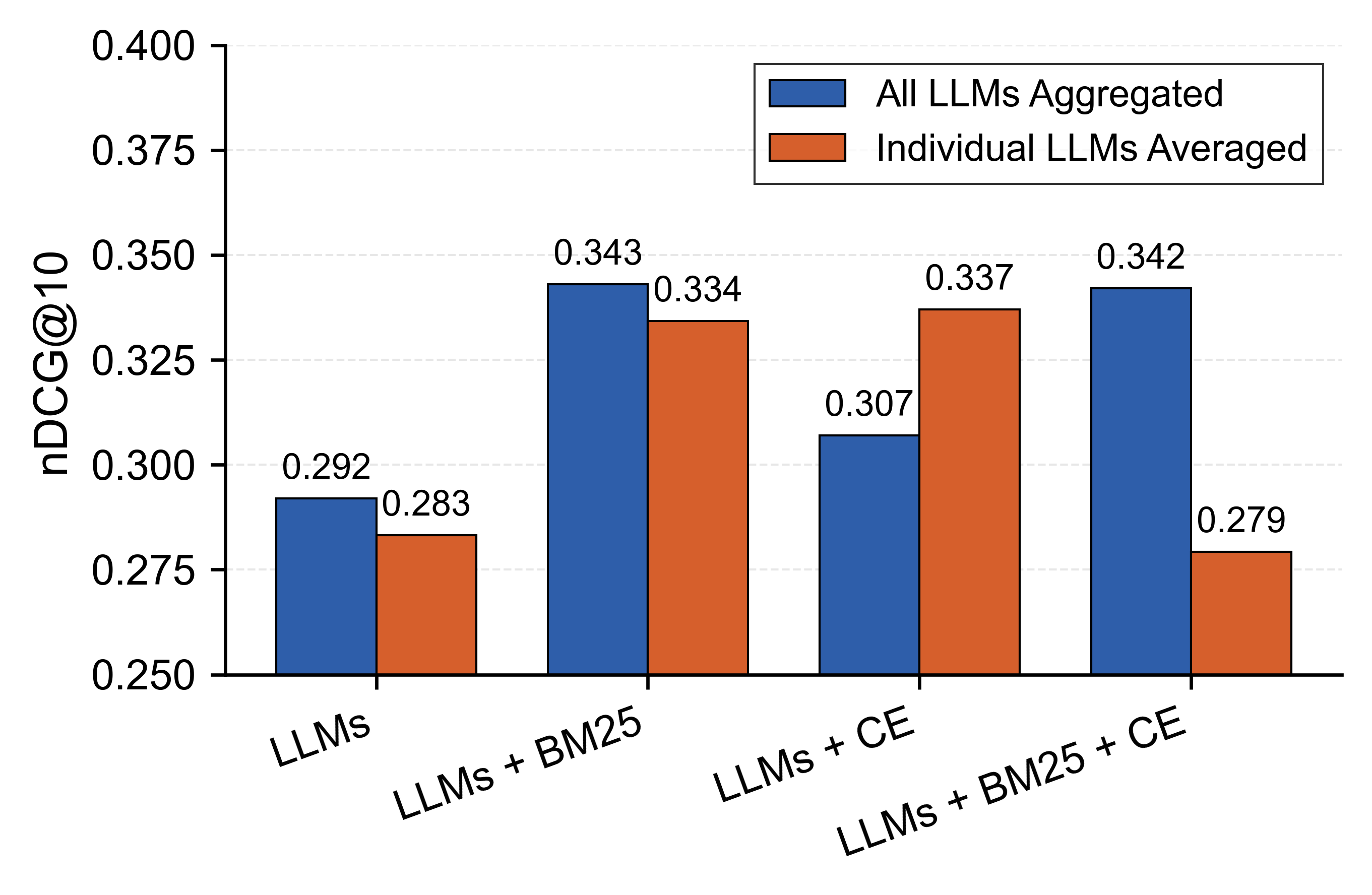}
        \caption{Average nDCG@10 across BEIR datasets comparing (i) all LLM-generated hard negatives aggregated, and (ii) performance averaged over individual LLMs, for vanilla LLM negatives, LLM + BM25, and LLM + cross-encoder (CE) mining.}
        \label{fig:concatenate}
    \end{subfigure}
    \caption{Performance analysis of DistilBERT fine-tuned on synthetic hard negatives.}
    \label{fig:main_overview}
\end{figure}

\paragraph{LLM Performance:}We generate hard negatives using four LLMs with sizes ranging from 4B to 30B parameters. As seen in Figure \ref{fig:line_plot} performance does not increase monotonically with model size. The best standalone LLM is Phi4-14B, achieving an average nDCG@10 of \textbf{0.314}, while the 30B model reaches \textbf{0.290}. Notably, the Qwen3-4B model achieves \textbf{0.282}, outperforming the 8B model (\textbf{0.247}). These results suggest that model scale alone does not determine the effectiveness of generated hard negatives.

\paragraph{Concatenation of Datasets:} We construct twelve datasets by naively concatenating LLM-generated hard negatives with each other and with BM25 and cross-encoder–mined hard negatives. As shown in Table\ref{tab:beir_results}, adding both BM25 and cross-encoder hard negatives provides a stronger training signal than LLM negatives alone, yet surprisingly, the traditional BM25 or cross-encoder supervision baselines still yield better retrieval performance overall. Notably, the LLaMA3-8B + CE dataset which incorporates negatives from the LLM that performed worst in isolation yields the second-best average performance within the CE + LLM configurations. Furthermore, fine-tuning DistilBERT on Phi4-14B + CE achieves the highest score on COVID(\textbf{0.357}) and Quora (\textbf{0.823}), surpassing the baselines, while the Qwen3-30B dataset leads to the lowest performance when concatenated with BM25 negatives. This underscores that larger LLMs do not necessarily generate more effective hard negative training signals.

\paragraph{Impact of Aggregation:} Figure\ref{fig:concatenate} compares the retrieval performance of a dense retriever fine-tuned on a dataset formed by concatenating hard negatives from all LLMs against the average performance of retrievers fine-tuned on individual LLM datasets. We observe that concatenating LLM-generated hard negatives leads to higher average performance than using individual LLM-generated datasets when combined with BM25, BM25+CE and the concatenated dataset as is; we also see that on NFCorpus(\textbf{0.235}) the ``All LLM" dataset is the best performing . Moreover, incorporating both BM25 and cross-encoder hard negatives substantially improves retrieval performance compared to dense retrievers trained on LLM-generated negatives alone. However, with the exception of LLaMA3-8B, we observe a performance degradation when combining signals: fine-tuning on the full combination of BM25, cross-encoder, and LLM hard negatives results in worse retrieval performance than fine-tuning on the standalone LLM-generated hard negatives.

\subsection{Analysis of Generated Hard-Negatives}

\begin{figure}[t]
    \centering

    \begin{subfigure}{0.32\linewidth}
        \centering
        \includegraphics[width=\linewidth]{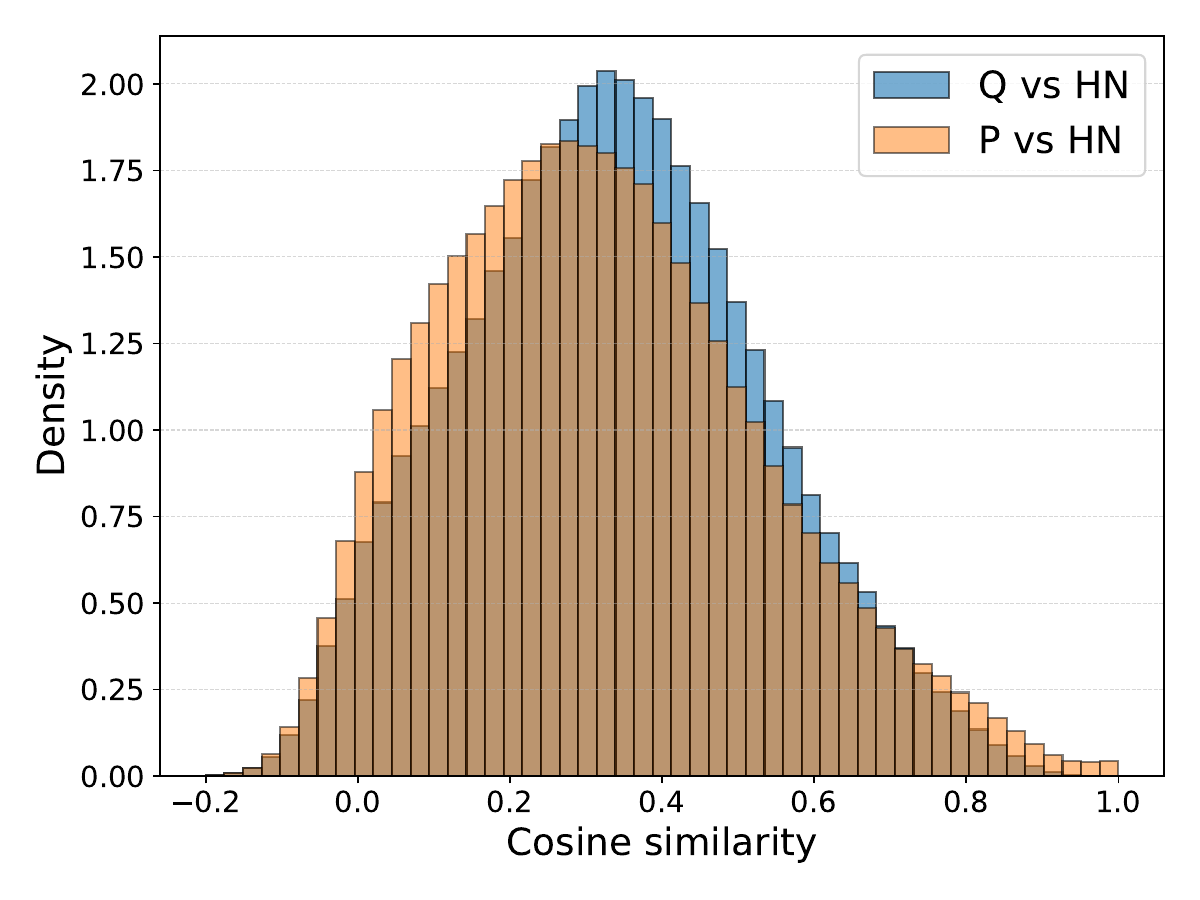}
        \caption{BM25}
    \end{subfigure}
    \hfill
    \begin{subfigure}{0.32\linewidth}
        \centering
        \includegraphics[width=\linewidth]{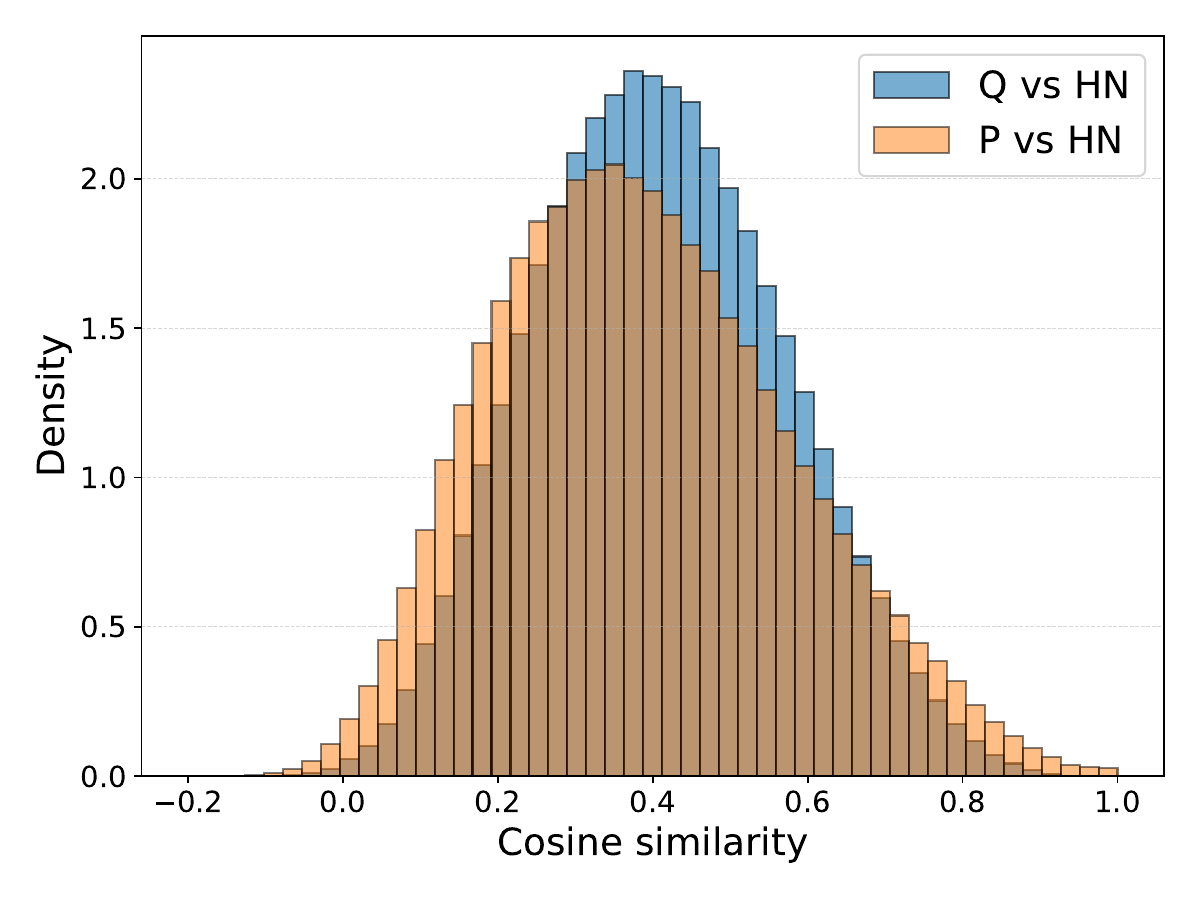}
        \caption{Cross-Encoder}
    \end{subfigure}
    \hfill
    \begin{subfigure}{0.32\linewidth}
        \centering
        \includegraphics[width=\linewidth]{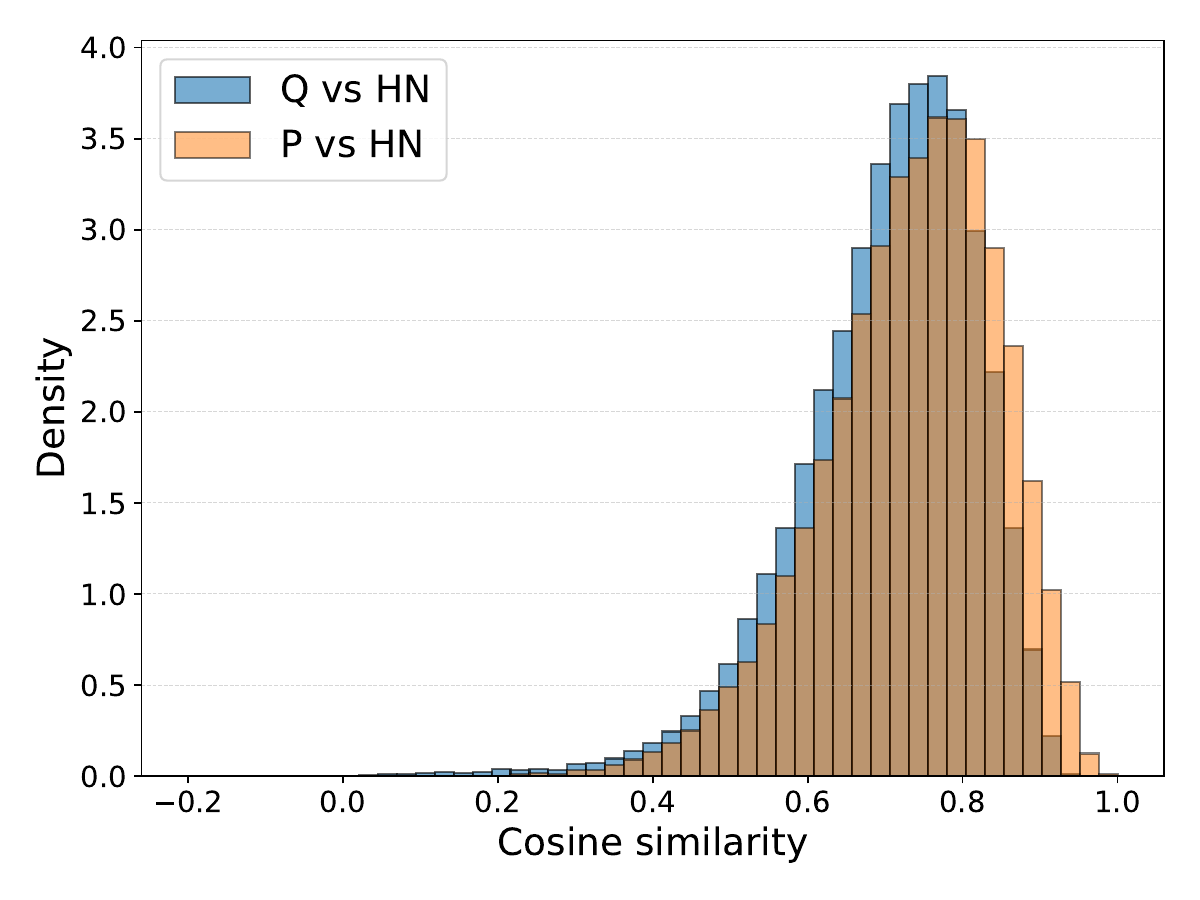}
        \caption{LLaMA3-8B}
    \end{subfigure}

    \vspace{0.5em}

    \begin{subfigure}{0.32\linewidth}
        \centering
        \includegraphics[width=\linewidth]{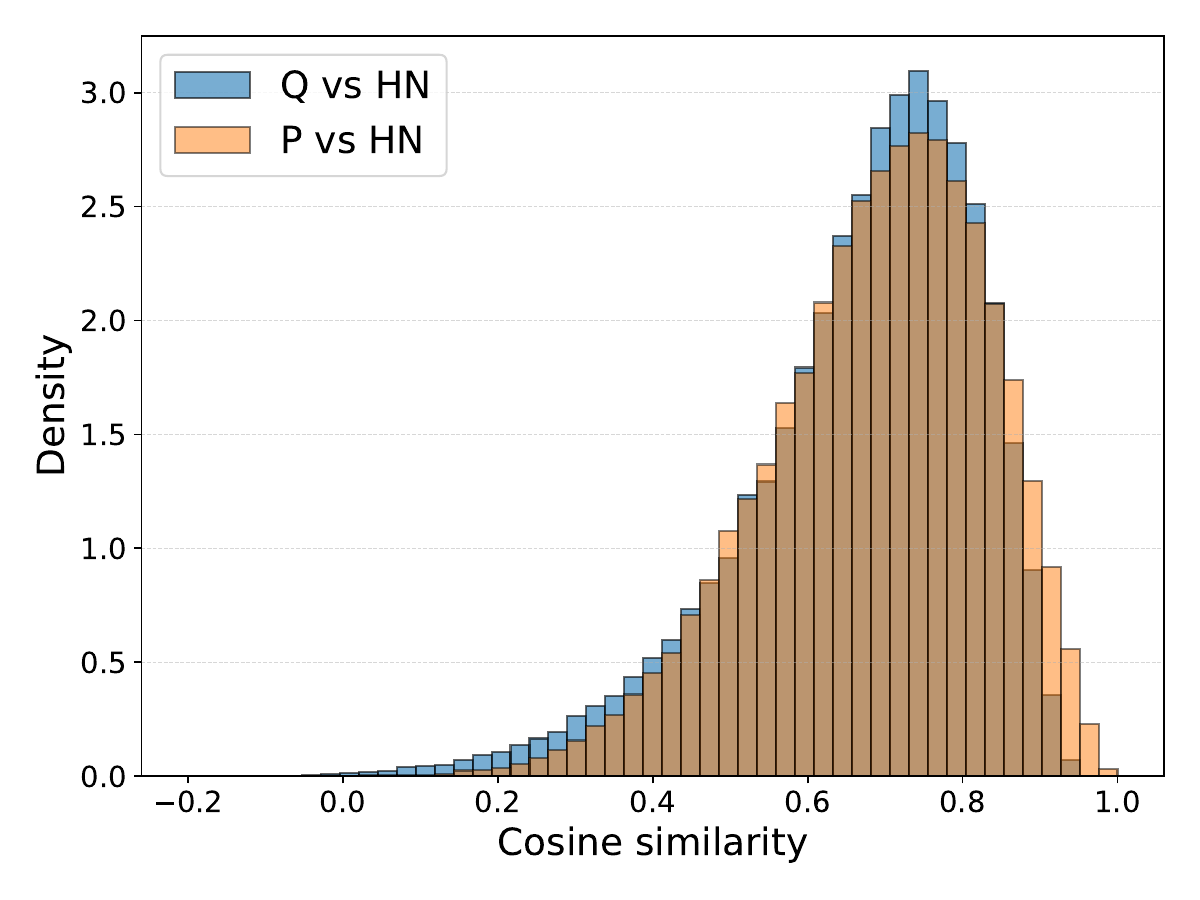}
        \caption{Phi-4-14B}
    \end{subfigure}
    \hfill
    \begin{subfigure}{0.32\linewidth}
        \centering
        \includegraphics[width=\linewidth]{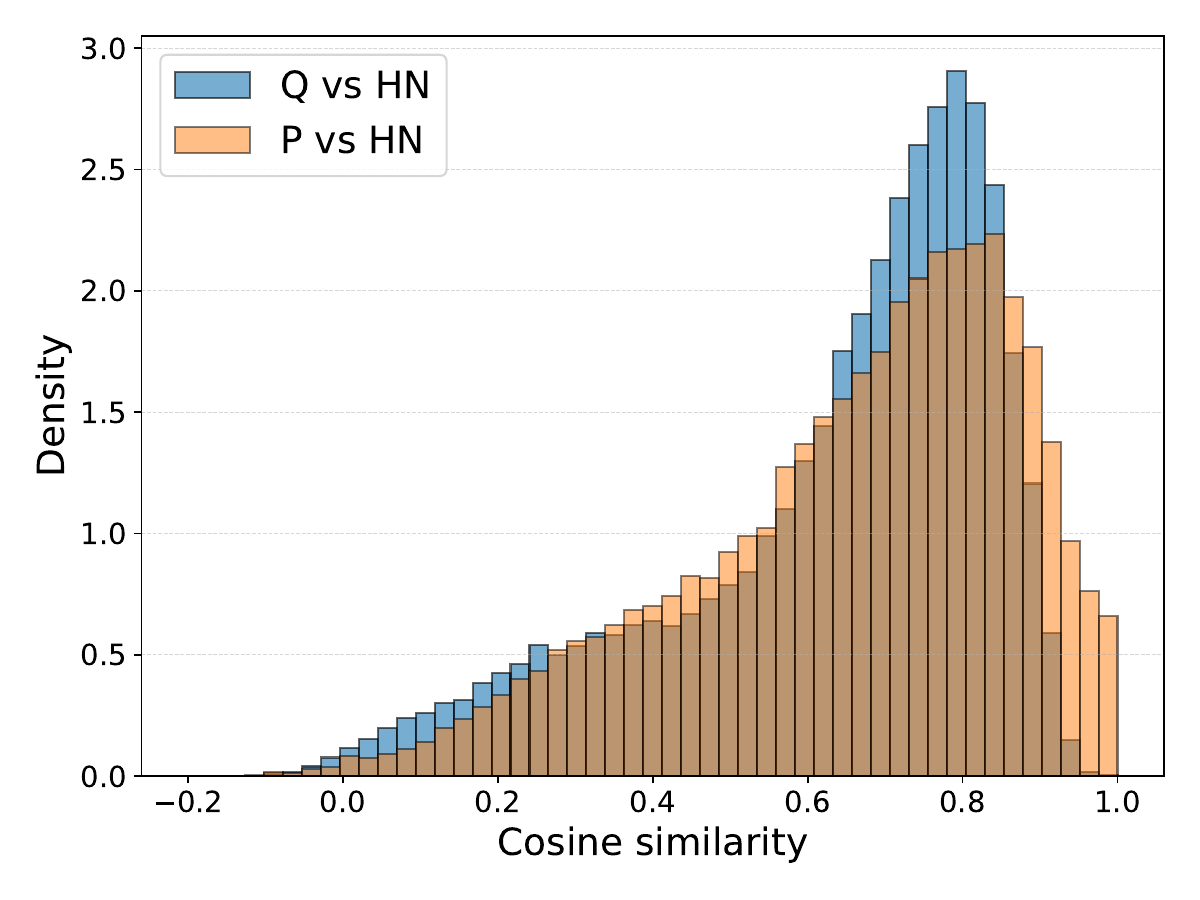}
        \caption{Qwen3-4B}
    \end{subfigure}
    \hfill
    \begin{subfigure}{0.32\linewidth}
        \centering
        \includegraphics[width=\linewidth]{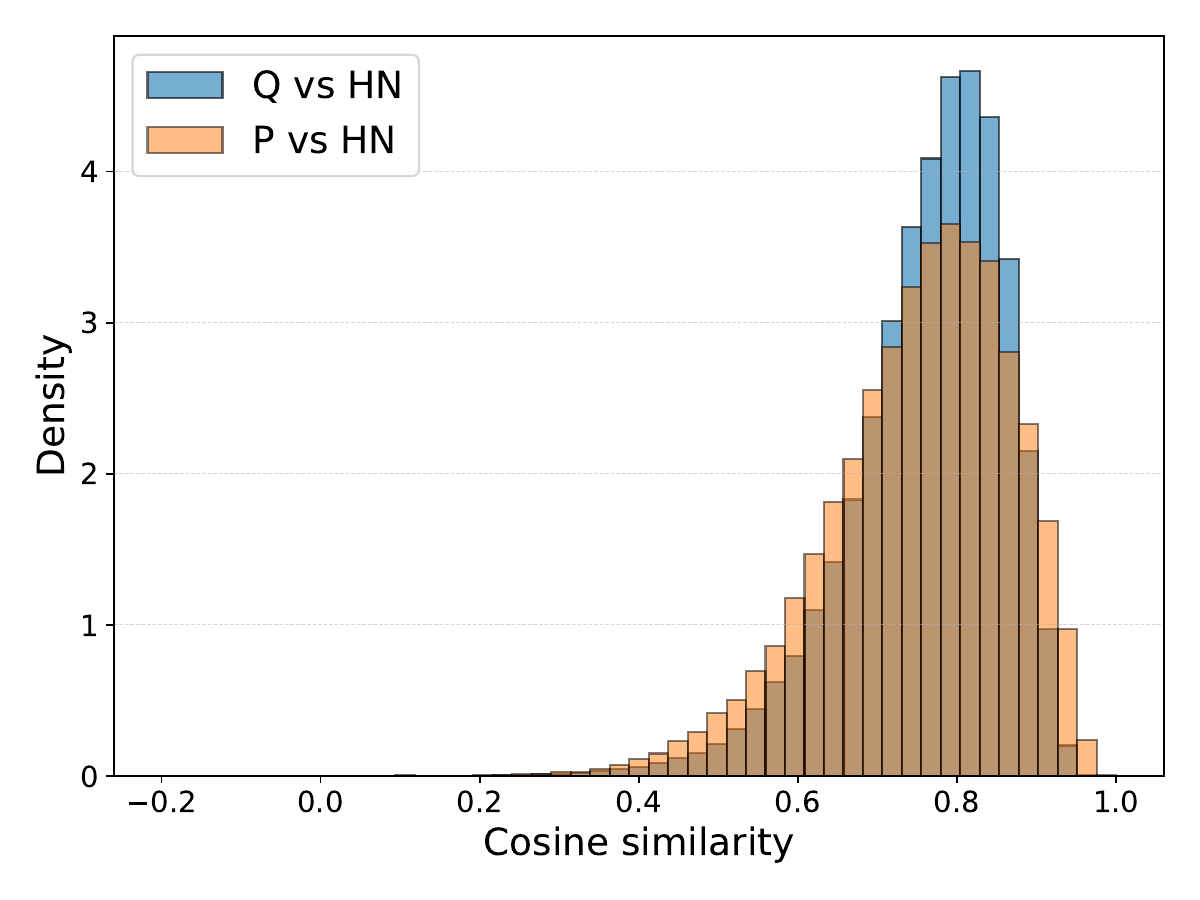}
        \caption{Qwen3-30B}
    \end{subfigure}

    \caption{Cosine similarity distribution of generated/mined hard negatives (HN) against the queries (Q) and positives (P).}
    \label{fig:hard_negative_cosine}
\end{figure}

In Figure \ref{fig:hard_negative_cosine}, we examine the cosine similarity distributions of the generated and mined hard negatives against the corresponding queries (Q) and positive documents (P). A consistent trend emerges across all LLM configurations: their similarity distributions are heavily concentrated near the upper bound (1.0). This implies that the LLMs are generating hard negatives that are overly similar to the ground-truth contexts, potentially bordering on false negatives. Conversely, the distributions for BM25 and cross-encoders span a significantly wider and more even range of similarities. These findings suggest that the superior performance of corpus-mined baselines stems from their ability to expose the retriever to a broader spectrum of negative signals, whereas a lack of variance in LLM-generated negatives limits the contrastive learning process.

\section{Conclusions}

We proposed a corpus-free pipeline for training dense retrievers by prompting LLMs to generate hard negatives, eliminating the need for full-corpus access during the mining phase. Our evaluation on the BEIR benchmark reveals that while LLM-generated negatives provide a viable training signal, they currently yield lower retrieval performance compared to traditional BM25 and cross-encoder mining strategies. Furthermore, we find that LLM scale is not a decisive factor in the quality of generated negatives, as smaller models such as Phi4-14B often outperformed larger counterparts like Qwen3-30B. Critically, our experiments with naive dataset concatenation demonstrate that simply combining synthetic and retrieved negatives can degrade performance or improve it. This inconsistency suggests that future research should prioritize developing advanced filtering or integration strategies to effectively leverage the complementary strengths of generative and retrieval-based supervision.

\bibliography{iclr2026_conference}
\bibliographystyle{iclr2026_conference}

\appendix
\section{LLM Configuration and Inference}\label{llm-config}
The model was loaded using \texttt{vllm} \cite{kwon2023efficient} for efficient inference, and all models were configured 
\begin{itemize}
    \item \textbf{Sampling parameters:}
    \begin{itemize}
        \item Temperature: $0.6$
        \item Top-$p$: $0.95$
        \item Top-$k$: $20$
        \item Minimum $p$: $0.0$
        \item Maximum tokens: $1024$
    \end{itemize}
    \item Tensor parallel size: $2$ \& $6$ (for the Qwen3-30B model)
    \item d\_type: \texttt{float32}
    \item GPU memory utilization: $0.80$
\end{itemize}

\section{Prompts}\label{prompts}
\subsubsection{User Prompt Template}
The user prompt template provides specific instructions for generating 5 passages, including length constraints and the required output format:

\begin{tcolorbox}[
  title=System Prompt for Hard Negative Generation,
  colback=gray!5,
  colframe=black,
  fonttitle=\bfseries,
]
\texttt{You are an assistant that generates hard negative passages for information retrieval tasks. A hard negative is a passage that seems relevant to the query but does not actually answer it or provide the correct information. You will be given both a query and a positive passage (the correct answer). Use this context to generate hard negatives that are semantically similar but factually incorrect or irrelevant.}
\end{tcolorbox}

\begin{tcolorbox}[
  title=User Prompt Template for Hard Negative Generation,
  colback=white,
  colframe=black,
  fonttitle=\bfseries
]
\textbf{Generate 5 hard negative passages for the following query and positive passage pair.}

The hard negatives should be similar in style and topic to the positive passage but should NOT correctly answer the query. Each passage should be max of 100 words but no less than 75.

\textbf{Query:} \{query\}

\textbf{Positive Passage (correct answer):} \{positive\}

Generate 5 hard negative passages that seem relevant but are actually incorrect or don't properly answer the query.

\textbf{Provide the passages in the following format:}

Passage 1: [your first passage]

Passage 2: [your second passage]

Passage 3: [your third passage]

Passage 4: [your fourth passage]

Passage 5: [your fifth passage]
\end{tcolorbox}

The \texttt{\{query\}} placeholder is dynamically replaced with the actual query for which negatives are being generated. The LLM's tokenizer was used to apply this chat template, ensuring the correct format for the model.

\end{document}